\documentclass[11pt]{article}

\usepackage[final]{acl}

\usepackage{times}
\usepackage{latexsym}

\usepackage[T1]{fontenc}

\usepackage[utf8]{inputenc}

\usepackage{microtype}

\usepackage{inconsolata}

\usepackage{graphicx}
\usepackage{amsmath}
\usepackage{booktabs}
\usepackage{multirow}
\usepackage{xcolor}
\usepackage{listings}
\usepackage{tcolorbox}

\newtcolorbox{promptbox}[1]{colback=white,colframe=black!60,boxrule=0.5pt,arc=2pt,left=4pt,right=4pt,top=3pt,bottom=3pt,title={#1},fonttitle=\footnotesize\bfseries,before skip=8pt,after skip=8pt}

\newcommand{\appendixsection}[1]{%
  \refstepcounter{section}%
  \section*{Appendix \thesection: #1}%
}

\title{LLM Post-Training as Brownfield Maintenance: An Industrial Perspective on Dataware Engineering}
\author{
  Gopi Krishnan Rajbahadur\textsuperscript{1} \quad
  Amir M. Ebrahimi\textsuperscript{1} \quad
  Boyuan Chen\textsuperscript{2} \quad
  Ahmed E. Hassan\textsuperscript{1} \\
  \textsuperscript{1}Queen's University, Canada \\
  \textsuperscript{2}Centre for Software Excellence, Huawei Technologies, Canada \\
  \texttt{grajbahadur@acm.org} \quad
  \texttt{amir.ebrahimi@cs.queensu.ca} \\
  \texttt{boyuan.chen1@huawei.com} \quad
  \texttt{ahmed@cs.queensu.ca}
}

\begin{document}
\maketitle

\begin{abstract}
Industrial post-training is a brownfield regime. Teams inherit a deployed checkpoint and must land targeted improvements under fixed compute and mixture budgets without regressing the rest. The maintained artifact is increasingly \textbf{dataware}: behavior governed by a curated post-training mixture, updated via bounded mixture patches rather than clean-slate retraining. From an industrial code-generation improvement effort, we offer a maintainer's perspective on why this work is hard in practice, distilling three recurring challenges, \textbf{zero-sum mixture design}, \textbf{yield as the binding metric}, and \textbf{end-to-end integration under uncertainty}, and arguing that progress depends less on one-off recipes than on an engineering discipline for programming dataware. In our case study, interventions that raised the conversion of teacher distillation into usable training data increased accepted supervision by \textbf{2.84$\times$} while using the same solution teacher and four solution attempts per candidate problem. In our primary evaluation, the yield-engineered patch improved CodeForces pass@1 by \textbf{+2.59} points (+3.11 pass@3) and held-out LiveCodeBench v6 pass@1 by \textbf{+6.11} (+8.05 pass@3), all statistically significant across 16 stochastic evaluations of each benchmark from one fixed checkpoint per condition, with internal AIME and MATH regression suites within tolerance.
\end{abstract}

\section{Introduction}
\label{sec:intro}

A familiar production mandate in industrial post-training is deceptively simple: \emph{improve a specific capability without degrading the rest}. In our case, improve code generation where our model struggled. We started from a brownfield baseline, an inherited checkpoint whose capabilities and failure modes could not be erased by starting fresh. For many teams, this is the everyday reality. 

The challenges and constraints are equally familiar and far less negotiable. First, the post-training mixture budget is fixed due to limited compute windows and shared cluster time, so any new data must earn its place by displacing existing data~\textbf{(Challenge C1: zero-sum mixture design)}~\citep{xie2023doremi,liu2025regmix}. Second, the data that would close the gap is scarce. For hard domains like competitive programming, high-quality human data is rare, so the obvious move is to distill from a stronger teacher. But distillation is costly, and the bottleneck is how much of what it produces is \emph{usable} after verification~\textbf{(Challenge C2: yield as the binding metric)}. Finally, even hard-won data is difficult to integrate into the mixture: it must lift the target capability without degrading the rest, yet capability losses surface as statistical shifts that take repeated, expensive evaluation to detect, and a local patch can induce non-local effects on global behavior~\textbf{(Challenge C3: end-to-end system integration)}.

Recent open recipes \citep{huggingfacetb2025smolplaybook} have made the greenfield creation phase increasingly accessible and well documented, but once a model is deployed the reality shifts from \emph{creation} to \emph{maintenance}. The flexibility of greenfield development disappears: teams cannot simply expand the mixture. These warnings are not new. A decade ago, Sculley et al.'s seminal NeurIPS paper \citep{sculley2015hidden} framed deployed ML as technical debt, and Brooks \citep{brooks1975mythical} had argued earlier still that systems fail when parts are optimized without end-to-end coherence; post-training inherits both, except the coupling now lives in data and evaluation rather than code.

\begin{figure}[t]
  \centering
  \includegraphics[width=\columnwidth]{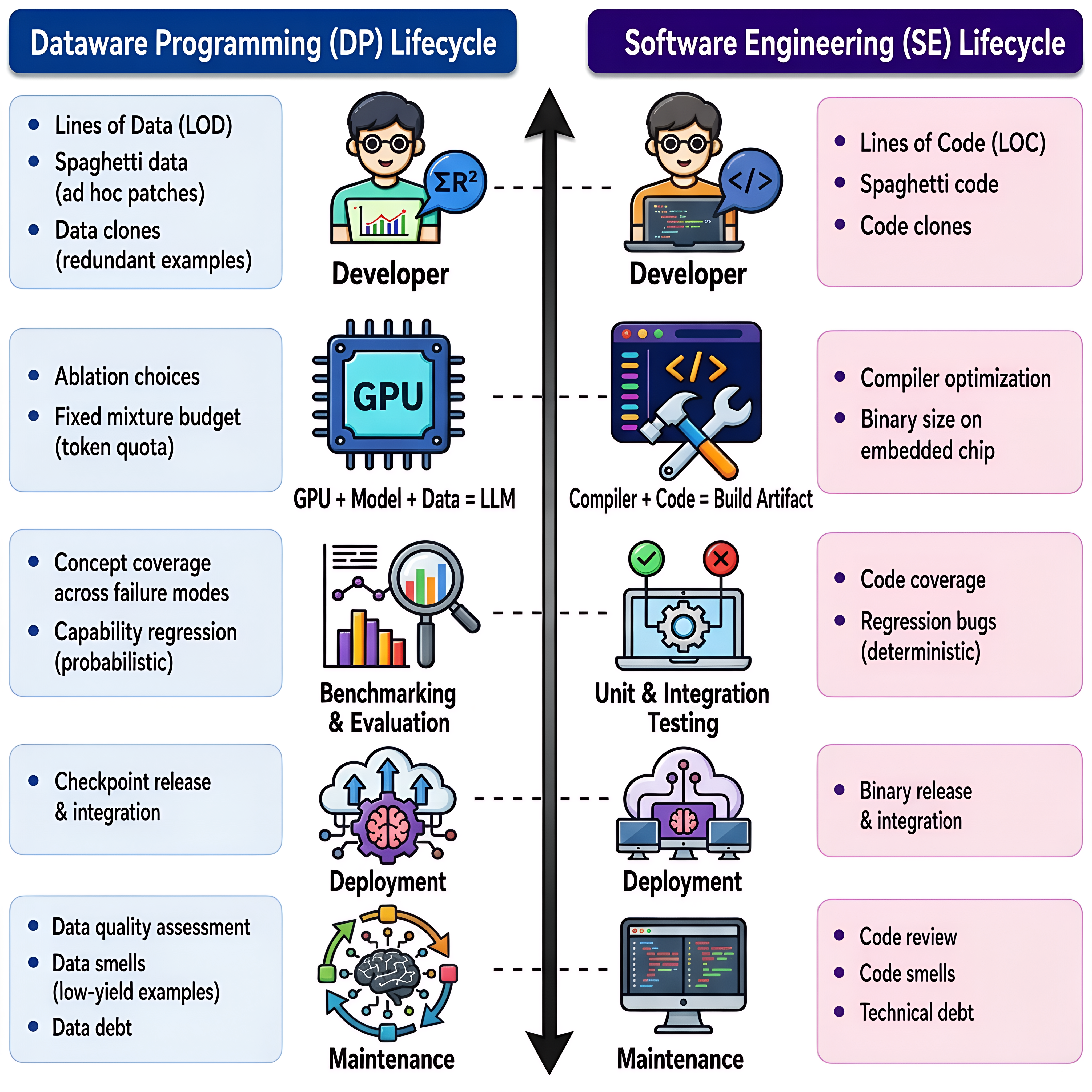}
  \caption{Structural parallels between brownfield software maintenance and brownfield LLM post-training.}
  \label{fig:structural-parallels}
\end{figure}

When the ``system'' is a large language model (LLM), much of the coupling and the ensuing complexity is encoded in the data itself, and post-training is \emph{programming with data}: each training example is an instruction, the mixture is the program, and the checkpoint is the compiled binary. In that sense, the maintained artifact is \textbf{dataware}, a system whose behavior is specified by a data program, the post-training mixture. Like mature software, it evolves by bounded patches rather than clean-slate rebuilds. This is \textbf{brownfield software maintenance} applied to an inherited checkpoint. Figure~\ref{fig:structural-parallels} summarizes the structural parallels (e.g., Lines of Code $\rightarrow$ Lines of Data), a maintainer's vocabulary for recurring pitfalls.

We document this regime from the maintainer's perspective and argue that treating data as code must become an operational contract, not a metaphor. From there, our paper makes three contributions. First, we frame industrial post-training as brownfield maintenance and distill the three coupled challenges (C1--C3) described above. Second, we report an industrial case study of failure-driven synthesis (FDS). Mining stable failures to guide synthetic data is established practice \citep{liang2025sws,li2025reversegen,cheng2024autodetect}. Our contribution is what adopting it in production demands, with outcomes validated by running one fixed checkpoint per condition on each benchmark 16 times, yielding 16 stochastic generations per task, and by reporting the operational signatures (yield, cost, coverage, stability) that decide feasibility. Third, we turn the recurring lessons into an agenda of five reviewable contracts for programming dataware and illustrate them with our case study.

\section{The Challenges of Data Programming}
\label{sec:challenges}
In the brownfield regime, post-training feels less like ``building a model'' and more like maintaining a mission-critical codebase on a release train. The community already has strong tools for synthesis, filtering, and mixture allocation \citep{wang2022selfinstruct,xie2023doremi,fan2024doge,li2025datamixingoptimizationsupervised}. The maintainer's pain is that these pieces interact in ways that are hard to anticipate and expensive to validate end-to-end. We take C1--C3 in turn, showing how each manifests in practice and why naive fixes fail under the constraints that bind in production.

\paragraph{Challenge C1: Zero-Sum Mixture Design}
The first hard limit is capacity. Token quotas, compute windows, and shared cluster time impose a fixed post-training mixture budget \citep{li2025datamixingoptimizationsupervised}. Every new example must \emph{earn its place} by displacing or reweighting existing content \citep{xie2023doremi,fan2024doge,liu2025regmix}. In maintenance, allocation becomes zero-sum: the model already encodes broad capabilities, so non-targeted additions yield diminishing returns while consuming scarce capacity, and the mixture quietly accumulates low-yield content as \emph{data debt}, echoing Lehman's laws of software evolution \citep{lehman1980programs}. This changes the first question from ``what data would help?'' to ``what data is worth \emph{displacing} to make room for it?''

\paragraph{Challenge C2: Yield as the Binding Metric}
Once capacity is fixed, what binds is not how much we can distill from the teacher but how much of it becomes \emph{usable} after verification. We use \textbf{yield engineering} to mean the deliberate design of \emph{upstream} and \emph{in-loop} controls that raise \textbf{distillation yield}, the number of valid, trainable solutions per unit \textbf{distillation effort} (teacher tokens, attempts). Within a fixed budget, these controls target avoidable failure modes in the data pipeline: underspecified or inconsistent problems, unhelpful long traces, and low-verifiability outputs. In our baseline (the case study of Section~\ref{sec:methodology}), distillation over 4,064 hard synthetic problems produced only 3,412 syntactically valid solutions (median valid solutions per problem: 0). When the teacher cannot solve a problem, effort produces cost but no signal, creating a \emph{yield trap}: a team can do everything right, good prompts, good filters, and still watch the usable fraction collapse, because generation cost scales linearly while signal plateaus \citep{ahmad2025opencodereasoning,he2025valleycodereasoningscaling}.

\paragraph{Challenge C3: End-to-End System Integration}
In brownfield post-training, the hard part is rarely proposing a change; it is integrating it safely. Regressions are stochastic: a mixture edit can improve one capability while degrading another, and evaluation noise can mask the damage. A single benchmark evaluation provides only one stochastic generation per task, so an apparent aggregate gain can depend on decoding noise. Section~\ref{sec:experiments} therefore runs one fixed checkpoint per condition on each benchmark 16 times, yielding 16 stochastic generations per task, and reports confidence intervals over tasks. This is the ``flight test'' problem: unlike deterministic unit tests, validating a change requires statistical judgment under variance \citep{madaan2024variance,rajput2025dynamicstability}. CodeForces-based evaluation is especially unstable across runs and contest selection~\citep{zheng2026whenelo}. In production, that variance becomes an operational question: under fixed budgets, how many runs are enough before we can trust that an improvement is real? Teams face a painful tradeoff, either replicate heavily and stall, or ship a change that wins once on a ``hero'' run and regresses later.

Compounding this, post-training inherits a combinatorial debugging surface. Pipelines expose many coupled decisions (data sources, templates, thresholds, quality gates, mixing weights) with non-linear interactions. Exhaustively ablating even five decision points with three options each already costs $3^{5}=243$ configurations before any replication. Under industrial budgets, ``just ablate it'' is off the table, so diagnosis degrades into best-guess trial and error.

The hardest failures appear at end-to-end integration. Components that look strong in isolation (a synthesis recipe, a filtering rule, a teacher model, mixing weights) can fail once combined under the same budget and regression constraints, exactly the loss of conceptual integrity Brooks warned against \citep{brooks1975mythical}. The symptom in post-training is non-local side effects: a local mixture edit that helps coding can destabilize unrelated behavior, because the update interacts with the inherited mixture and learned representations in hard-to-predict ways \citep{wang2024inscl}. Maintainers then stop asking ``is this dataset good?'' and ask ``will it compose with what the model already is?''

\paragraph{Summary.}
These challenges are coupled: fixed budgets (C1) make yield the binding constraint (C2), and yield-focused changes amplify integration risk (C3).

\section{A Maintenance Industrial Example}
\label{sec:methodology}
The Failure-Driven Synthesis (FDS) recipe is standard in both industry and academia: mine a model's stable failures, extract the concepts behind them, and synthesize targeted training data. The same pattern recurs across settings: reinforcement learning (RL) for mathematical reasoning with the PromptCoT-style synthesis we also use (SwS; \citealp{liang2025sws}); a trained proposer that surfaces failure-inducing queries for safety, honesty, and math (ReverseGen; \citealp{li2025reversegen}); distillation of 553K instruction-tuning examples from the reasoning failures of multimodal models \citep{stan2026learningreasoningfailuressynthetic}; automated weakness detection feeding targeted improvement data (AutoDetect; \citealp{cheng2024autodetect}); and synthesis around failure regions to steer second-stage fine-tuning and RL (Logics-STEM; \citealp{xu2026logicsstem}). We claim no novelty for it. What these works leave undocumented is what it takes to run the recipe in an industrial setting, where the three challenges of Section~\ref{sec:challenges} bind at once. Prior work typically reports the gains; we document what it costs to reach them under production constraints. 

\subsection{Setting: a bounded patch on a live system (C1, C3)}
\label{sec:setting}

We start from \textbf{TuringThinker-7B} ($\mathcal{M}_0$), an internal general-purpose, reasoning-capable checkpoint, and target improved competitive-programming performance within a fixed training budget. $\mathcal{M}_0$ is a deployed checkpoint, so capability improvements ship on a release cadence gated by an internal regression suite. Its post-training mixture is capacity-bounded (C1): we cannot disclose the exact token budget, but the code-related allocation was roughly 650K training examples (Lines of Data, or LOD). Because any change must integrate without degrading global behavior (C3), the FDS patch is a bounded modification to the post-training ``data program'' (Figure~\ref{fig:structural-parallels}): it may displace at most 2\% of the 650K, and the full patch shipped in production contains 9,697 examples, about 1.5\% of the code allocation. For controlled evaluation, Section~\ref{sec:setup} uses a 3,412-example coverage-first subset, FDS-3K-Cov. In production, a capability patch enters the mixture only by displacing matched-volume, lower-quality synthetic code data, with reweighting confined to the code allocation, and only if it clears a strict acceptance gate: it must improve the target benchmarks (CodeForces and LiveCodeBench v6) while AIME and MATH, our internal regression suite, fall by no more than one point. This protects existing math and general-reasoning capabilities while we patch the model's code.

\subsection{From stable failures to a yield-positive data patch}
\label{sec:pipeline}
\begin{figure*}[t]
  \centering
  \includegraphics[width=0.75\textwidth]{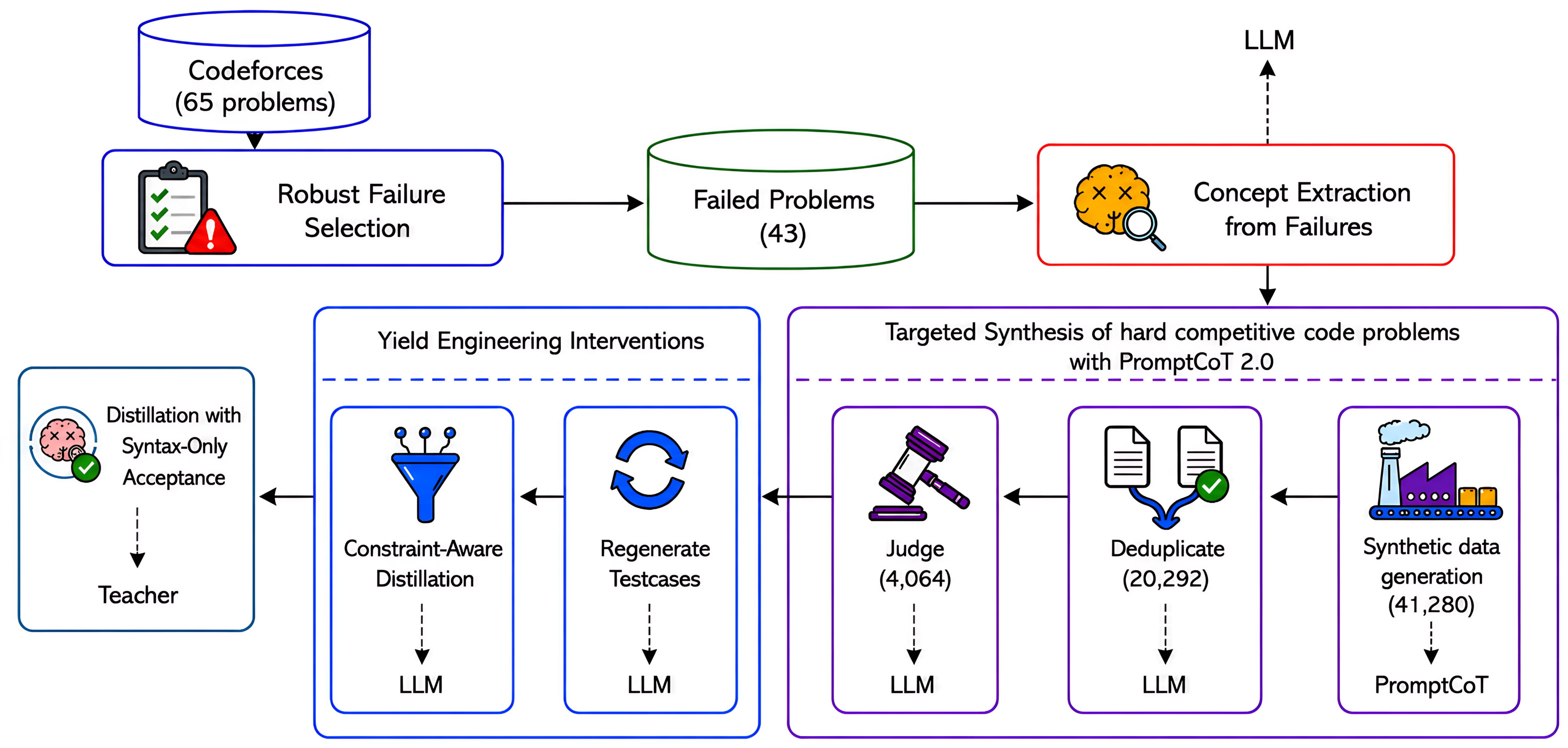}
  \caption{FDS pipeline. Stage tags: stable-failure identification (C3), concept-conditioned synthesis and quality gates (C1), yield-optimized distillation (C2). Funnel: 41,280 generated $\rightarrow$ 20,292 deduplicated $\rightarrow$ 4,064 judge-approved.}
  \label{fig:revisited-pipeline}
\end{figure*}

FDS turns reproducible failures into a targeted patch that fits within a fixed mixture budget. Figure~\ref{fig:revisited-pipeline} shows the pipeline, with its C1--C3 stage tags and funnel counts.
\paragraph{Stability first: deciding what is worth fixing (C3).}
Single-pass evaluation is unreliable under stochastic generation, so before synthesizing anything we narrowed the target to failures we could trust by defining the failure set $\mathcal{F}$ as the intersection across repeated evaluations: a task enters $\mathcal{F}$ only if none of its $N$ samples passes in any of the $R$ mining runs, $\mathcal{F}=\bigcap_{r=1}^{R}\{q \mid \text{pass@}N(q, \mathcal{M}_0)_r = 0\}$. We use $R=3$ and $N=3$ for mining, and separately run each fixed checkpoint on each benchmark 16 times, yielding 16 stochastic generations per task for the variance-aware reporting of Section~\ref{sec:experiments}. This deliberately prioritizes precision over recall: under a fixed budget, we prefer to spend synthesis and distillation effort on durable weaknesses rather than chase evaluation noise.
\paragraph{Structure next: keeping patches coherent (C3).}
A naive response to $\mathcal{F}$ is to synthesize ``more hard problems'', but in practice that quickly becomes unstructured accumulation that integrates poorly with the base model (C3). To preserve conceptual integrity, we routed synthesis through a small concept taxonomy extracted from the model's own failure modes: an LLM tagger (\textbf{DeepSeek-V3-1-Terminus}, prompt in Appendix~\ref{sec:appendix-prompts}) labeled each failure in $\mathcal{F}$ with curriculum-level concepts (e.g., \emph{Dynamic Programming}, \emph{Two Pointers}), keeping the five it judged most essential to solving each problem. We then generated concept-conditioned problems with \textbf{PromptCoT 2.0} \citep{zhao2025promptcot2}. Because the distillation budget is the binding cost, we moved quality checks left: we semantically deduplicated the candidates and applied an LLM judge (\textbf{GLM-4.6}, Appendix~\ref{sec:appendix-prompts}) to keep only coherent, solvable problems, the pool eligible for distillation (Figure~\ref{fig:revisited-pipeline}).
\paragraph{Yield finally: two interventions to escape the yield trap (C2).}
Even after filtering, distillation exposed the real bottleneck: on hard synthetic problems, teachers frequently fail to produce valid solutions, so distillation cost grows faster than the distillation yield resulting in a yield trap (C2). In practice, two failure modes dominated, and each forced an explicit intervention; neither is novel on its own.

\textbf{Intervention 1: \emph{Testcase Rectification}.}
We observed \emph{specification misalignment}, where example testcases contradicted the statement. We therefore regenerate representative I/O before distillation, a standard data-cleaning step.

\textbf{Intervention 2: \emph{Constraint Injection}.}
We observed \emph{inefficient reasoning}, where the teacher spent tokens re-checking constraints without producing runnable code. We therefore prepend the extracted constraints to focus its reasoning, an instance of Thinking Intervention \citep{wu2025thinkingintervention}.

Neither intervention is novel, but both are seldom reported as part of published failure-driven-synthesis pipelines, an omission that leaves teams in the yield trap. Both serve one objective: improving distillation yield. 

\paragraph{Keeping the loop affordable (C1, C2, C3).}
Under fixed budgets (C1) and a combinatorial intervention surface (C3), execution-based verification inside the loop was not feasible. We instead allowed at most four teacher queries per synthetic problem, in both the standard and FDS settings, and gated admission with a deterministic Tree-sitter check \citep{tree_sitter}: a response is kept only if it parses as valid Python. The check is syntactic, not semantic, so it cannot guarantee correctness. It is purely a cost-control gate that raises yield per teacher token (C2).

\section{Experimental Setup and Results}
\label{sec:experiments}

\begin{table}[t]
  \centering
  \scriptsize
  \setlength{\tabcolsep}{3pt}
  \begin{tabular}{@{} l l p{0.54\linewidth} @{}}
    \toprule
    \textbf{Signature} & \textbf{Challenge(s)} & \textbf{Evidence} \\
    \midrule
    Budget & \textbf{C1} &
    Mean teacher tokens/attempt \textbf{$-28\%$} (20{,}498 $\rightarrow$ 14{,}683); median trace length \textbf{$-46.5\%$} (20{,}634 $\rightarrow$ 11{,}048) at the same four-attempt cap (Figure~\ref{fig:reasoning-length}). \\

    Yield & \textbf{C2} &
    Usable supervision \textbf{$\times 2.84$} at fixed attempts: 3{,}412 $\rightarrow$ 9{,}697 accepted solutions. \\

    Coverage & \textbf{C2}, \textbf{C3} &
    Non-zero problem coverage 1{,}044 $\rightarrow$ 2{,}699; \textbf{1{,}695 (63\%)} were zero-yield under baseline. \\

    Stability & \textbf{C3} &
    Each fixed checkpoint is evaluated 16 times per benchmark. Mixture-replacement FDS-3K-Cov pass@1: CodeForces \textbf{15.19} [8.75, 22.40]; LCBv6 \textbf{48.75} [42.61, 54.86] (95\% task-bootstrap CIs). \\

    Integration & \textbf{C1}, \textbf{C3} &
    Patch \textbf{= 3.5\%} of the continued-training mixture by example count (9{,}697 of 278{,}895); in deployment it enters by displacing matched-volume, lower-quality synthetic code within the code allocation (Sec.~\ref{sec:methodology}). \\
    \bottomrule
  \end{tabular}
  \caption{Operational signatures of the FDS case study.}
  \label{tab:episode-signatures}
\end{table}

\subsection{Setup}
\label{sec:setup}

We evaluate five continued-training conditions from the same \textbf{TuringThinker-7B} checkpoint, organized into three setups: Base, Additive, and Mixture replacement (Table~\ref{tab:benchmark_results}).

Mixture replacement is our primary brownfield setting: a capability patch must fit within a capacity-bounded training mixture and therefore displace existing data. The additive conditions provide a less constrained reference, testing whether the synthesized supervision remains useful when mixture capacity is relaxed and no rehearsal examples need be removed. Comparing the two setups therefore separates the value of the new supervision from the difficulty of integrating it under a fixed mixture budget.

\noindent\textbf{Base.} The Base (no patch) condition trains on the inherited 269,198-example rehearsal mixture alone, controlling for the continued-training step.

\noindent\textbf{Additive.} We append one of two evaluation patches to the same rehearsal mixture: Distill-3K contains 3,412 solutions from standard distillation, and FDS-3K-Cov contains 3,412 coverage-first examples from the yield-engineered corpus. We construct FDS-3K-Cov without inspecting benchmark outcomes by retaining at least one example for each of the 2,699 FDS task IDs and filling the remaining 713 slots uniformly at random. It matches Distill-3K in added examples and total example count, but not in content, task distribution, or token count. For a fair comparison with Distill-3K, Table~\ref{tab:benchmark_results} uses FDS-3K-Cov.

\noindent\textbf{Mixture replacement.} We insert either Distill-3K or FDS-3K-Cov while removing 3,412 rehearsal examples, keeping the total mixture size fixed by example count. Displacement is restricted to duplicate records and alternate responses whose prompts remain represented, so no prompt is removed entirely.

The patch-generation comparison is \textbf{solution-teacher-attempt-matched}: both pipelines use the same solution teacher, the same four-attempt cap per candidate problem, and fixed solution-decoding settings. This budget excludes the auxiliary calls for testcase rectification and constraint extraction; FDS therefore evaluates the complete yield-engineering package rather than the causal effect of an individual intervention. The auxiliary models (tagger, generator, judge) are named in Section~\ref{sec:pipeline}, and the models used for testcase regeneration and constraint extraction are named in Appendix B.5. Student training uses identical hyperparameters across conditions (32K context, batch size 64, 6 epochs), with no condition-specific hyperparameter search.

We evaluate on \textbf{CodeForces} (65 tasks; \citealp{zheng2026whenelo}) and \textbf{LiveCodeBench v6} (LCBv6; 175 tasks; \citealp{jain2024livecodebench}). For each condition, we run one fixed checkpoint on each benchmark 16 times under fixed decoding settings, obtaining $n{=}16$ stochastic generations per task. Following standard code-generation evaluation \citep{chen2021evaluating}, for a task $q$ with $c_q$ execution-correct generations we compute $\widehat{\mathrm{pass@}k}(q)=1-\binom{n-c_q}{k}/\binom{n}{k}$ and average the task-level estimates across the benchmark; we report $k\in\{1,3\}$. Pass@1 measures single-sample reliability, while pass@3 estimates the probability that at least one solution succeeds under a three-generation budget. Following variance-aware evaluation practice \citep{miller2024errorbars,madaan2024variance}, we obtain 95\% confidence intervals by nonparametric bootstrap over benchmark tasks. For comparisons against Base, we bootstrap paired task-level differences and regard a difference as statistically significant when its 95\% interval excludes zero. All training data is decontaminated against both benchmarks' problem statements and hidden tests using $n$-gram overlap and embedding-similarity matching. Full harness details are in Appendix~\ref{app:benchmarks}.

\subsection{Results}
\label{sec:results}

\begin{table*}[t]
\raggedright
\footnotesize
\setlength{\tabcolsep}{1.55pt}
\renewcommand{\arraystretch}{1.2}
\begin{tabular}{@{\hspace{2pt}}c@{\hspace{8pt}}l ll ll ll @{}}
\toprule
\multirow{2}{*}{\textbf{Setup}} & \multirow{2}{*}{\textbf{Condition}} & \multicolumn{2}{c}{\textbf{CodeForces}} & \multicolumn{2}{c}{\textbf{LCBv6 (held-out)}} & \multicolumn{2}{c}{\textbf{Regression (avg@3)}} \\
\cmidrule(lr){3-4} \cmidrule(lr){5-6} \cmidrule(l){7-8}
 & & \multicolumn{1}{c}{pass@1} & \multicolumn{1}{c}{pass@3} & \multicolumn{1}{c}{pass@1} & \multicolumn{1}{c}{pass@3} & \multicolumn{1}{c}{AIME} & \multicolumn{1}{c}{MATH} \\
\midrule
\multirow[c]{1}{*}{\scriptsize Base} & Base (no patch) & 12.60 {\scriptsize[6.92, 18.94]} & 20.66 {\scriptsize[12.46, 29.56]} & 42.64 {\scriptsize[36.36, 48.93]} & 52.67 {\scriptsize[45.86, 59.45]} & 53.33 & 93.33 \\
\midrule
\multirow[c]{2}{*}{\scriptsize Additive} & Distill-3K & 12.88 {\scriptsize[6.92, 19.62]} & 20.52 {\scriptsize[12.45, 29.34]} & 43.75 {\scriptsize[37.46, 50.11]} & 53.80 {\scriptsize[46.97, 60.67]} & 61.67 & \textbf{93.87} \\
 & FDS-3K-Cov & 13.65 {\scriptsize[7.69, 20.29]} & 22.26 {\scriptsize[13.98, 31.28]} & \textbf{49.39*} {\scriptsize\textbf{[43.25, 55.50]}} & \textbf{61.55*} {\scriptsize\textbf{[55.11, 67.88]}} & 59.44 & 92.40 \\
\midrule
\multirow[c]{2}{*}{\scriptsize\shortstack[c]{Mixture\\replacement}} & Distill-3K & 14.90* {\scriptsize[8.75, 21.83]} & \textbf{24.36*} {\scriptsize\textbf{[15.88, 33.49]}} & 47.18* {\scriptsize[40.75, 53.57]} & 56.96* {\scriptsize[50.12, 63.72]} & \textbf{63.89} & 93.33 \\
 & FDS-3K-Cov & \textbf{15.19*} {\scriptsize\textbf{[8.75, 22.40]}} & 23.77* {\scriptsize[15.04, 33.05]} & 48.75* {\scriptsize[42.61, 54.86]} & 60.72* {\scriptsize[54.22, 67.11]} & 57.78 & 93.00 \\
\bottomrule
\multicolumn{8}{@{}p{\linewidth}@{}}{\scriptsize * paired 95\% bootstrap interval for the difference vs.\ Base excludes zero. \textbf{Bold} = best result in column.} \\
\end{tabular}
\caption{Downstream performance across the Base, additive, and mixture-replacement setups. We run one trained checkpoint per condition on each benchmark 16 times, yielding $n{=}16$ stochastic generations per task. Values are percentages; brackets give 95\% nonparametric bootstrap confidence intervals over benchmark tasks (20,000 replicates).}
\label{tab:benchmark_results}
\end{table*}

Table~\ref{tab:benchmark_results} shows that \textbf{yield engineering makes the targeted supervision substantially more useful across both studied setups}. In the additive setup, standard Distill-3K remains close to Base: its changes range from $-0.14$ to $+1.13$ points across CodeForces and LCBv6, with no statistically significant improvement. In contrast, FDS-3K-Cov improves all four target metrics: CodeForces pass@1 and pass@3 increase by $+1.05$ and $+1.60$ points, while LCBv6 pass@1 and pass@3 increase by $+6.75$ and $+8.88$. The LCBv6 gains are statistically significant. Thus, when mixture capacity is relaxed, the yield-engineered corpus provides substantially more effective supervision than the same-sized standard-distillation patch.

The same advantage largely persists in our primary brownfield setting, mixture replacement. Both patches significantly outperform Base on all four target metrics, showing that targeted supervision can survive displacement of rehearsal data. However, FDS-3K-Cov provides the strongest overall target-benchmark profile: it improves CodeForces pass@1 by $+2.59$ points and pass@3 by $+3.11$, and LCBv6 pass@1 and pass@3 by $+6.11$ and $+8.05$. Relative to Distill-3K, FDS-3K-Cov is higher on three of the four target metrics, including both held-out LCBv6 measures; Distill-3K is higher only on CodeForces pass@3 ($+3.70$ versus $+3.11$). These results make the yield-engineered corpus the more consistently useful patch across the additive and capacity-constrained settings.

Importantly, this improvement is not confined to the failures used to construct the patch. Failure mining used only the 65 CodeForces tasks, whereas LCBv6 was never used for mining and serves as the held-out transfer benchmark. The substantially larger LCBv6 gains from FDS-3K-Cov therefore indicate transfer beyond the specific failures that drove synthesis. The gains are also concentrated in targeted areas rather than appearing as a uniform robustness shift: concept-level improvement increases with targeted synthetic volume (Appendix Figure~\ref{fig:concept-volume}), which we treat as diagnostic rather than causal because mixture effects remain coupled.

Crucially, the stronger downstream utility does not come at the expense of the deployment gate or distillation efficiency. Under mixture replacement, FDS-3K-Cov raises AIME from 53.33 to 57.78 while changing MATH from 93.33 to 93.00; across all displayed conditions, regression-suite changes remain within the one-point tolerance. At the same time, the yield-engineering interventions increase accepted supervision from 3,412 to 9,697 solutions ($2.84\times$) under the same four-attempt cap, while reducing mean teacher tokens per attempt by $28\%$ and median trace length by $46.5\%$. The corpus that is more consistently useful downstream is therefore also materially more efficient to produce.

\section{Toward a Discipline of Programming Dataware}
\label{sec:discipline}

\paragraph{From industrial example to discipline.}
Software engineering matured by turning ad-hoc practice into shared methods and a culture of reliability, and post-training is at the same inflection point: open playbooks share \emph{creation} practice \citep{huggingfacetb2025smolplaybook}, while \emph{maintenance} contracts and tooling are still missing \citep{sculley2015hidden,madaan2024variance}. Frontier labs are converging on the same inversion: a recent Microsoft AI report frames its goal as a hill-climbing machine that outlasts any single model \citep{microsoftai2026maithinking}, but for the greenfield case. We document the maintenance loop after the model ships.

\paragraph{Essential vs.\ accidental complexity.}
Following Brooks' distinction, \textbf{essential complexity} here includes probabilistic outputs, non-local interactions, and a continuous, high-dimensional capability surface; these demand fundamental research. \textbf{Accidental complexity} comes from our tools and norms: we rarely measure how much teacher distillation becomes usable supervision, change mixtures without accounting for displacement, or integrate patches without explicit acceptance criteria. In our example the dominant costs were accidental: yield collapse made naive distillation uneconomic, and variance made single-run iteration non-actionable (Table~\ref{tab:episode-signatures}).

\paragraph{An agenda framed as engineering contracts.}
Five recurring lessons can become explicit, reviewable contracts, each tagged with the challenge it addresses.

\paragraph{(1) Per-LOD valuation (\textbf{C1}).}
We need scalable estimates of a patch's marginal value within a mixture, so teams can justify its budget share and negotiate trade-offs explicitly rather than implicitly.

\paragraph{(2) Probabilistic regression frameworks (\textbf{C3}).}
Regression is statistical and rarely localized: a single stochastic benchmark evaluation can produce a noisy aggregate, so single-evaluation deltas are not actionable. We need variance-aware acceptance criteria and regression gates calibrated to real evaluation regimes. In our experiments, each fixed checkpoint is evaluated 16 times per benchmark, and Table~\ref{tab:benchmark_results} reports task-bootstrap intervals.

\paragraph{(3) Experimental design for programming dataware (\textbf{C1}, \textbf{C3}).}
When interactions make diagnosis hard and exhaustive ablation is unaffordable, we need information-efficient experiment design: sequential planning, fractional-factorial thinking, and proxy gates that move validation earlier.

\paragraph{(4) Integration expectations and displacement accounting (\textbf{C1}, \textbf{C3}).}
Mixture components need explicit expectations: what they target, what budget they consume, what they displace, and what regressions they risk. Patch sizing, displacement baselines, and conservative merges should be shared practice, not ad-hoc craft.

\paragraph{(5) Reporting what makes progress feasible (\textbf{C2}, \textbf{C3}).}
The data program is the living part of the \textit{dataware}, so reporting only downstream scores hides what determines feasibility: yield per teacher token, cost per accepted sample, gate-survival rates, and return on distillation effort. In our example these signatures (Table~\ref{tab:episode-signatures}, Figure~\ref{fig:revisited-pipeline}), not the benchmark deltas, were what made improvement possible.

\paragraph{From art to engineering.}
The goal is not to eliminate intuition but to codify what works, so we stop paying for it at training-run prices. As models grow more consequential and more expensive, the field needs shared contracts, tools, and protocols that make mixture changes reviewable, testable, and safely integrable.

\section{Conclusion}
\label{sec:conclusion}
Post-training under fixed budgets is brownfield software maintenance: a deployed model improved by budgeted, regression-safe patches, not clean-slate rebuilds. None of the techniques we adopted is new, and that is the point: the open problem is the gap between research and practice, the constraints of adopting published methods safely on a live system. We offer this case study as evidence, not prescription, toward closing that gap: \textit{a discipline of programming dataware}.
\section*{Limitations}
\label{sec:limitations}

This is an industry perspective paper grounded in one case study: a single internal checkpoint family (TuringThinker-7B), one domain (competitive programming), and two benchmarks of 65 and 175 tasks. We do not provide full factorial ablations over all intervention dimensions; this is partly a direct consequence of the ablation-explosion constraint discussed in Section~\ref{sec:challenges}. Some implementation details, such as the exact mixture composition, the teacher model's identity, the actual token cap, and other internal measurements, cannot be fully disclosed due to the competitive nature of this setting; the core message is not the technique or the exact curves but the operating regime and the engineering constraints that bind.

In Table~\ref{tab:benchmark_results}, Distill-3K and FDS-3K-Cov are matched by added example count (3,412 each), but not by content, task distribution, or token count. FDS evaluates the complete yield-engineering package rather than an isolated intervention, so the comparison does not disentangle coverage, data composition, and other pipeline effects. Attribution is also inherently partial because FDS data is mixed with a larger rehearsal mixture during continued training. The full 9,697-example FDS-9K patch was shipped in production but is omitted from the controlled benchmark table.

Our distillation acceptance gate is syntax-only (complete Tree-sitter-valid code). This can admit semantically incorrect programs, so yield should be interpreted as an operational proxy for usable supervision rather than direct correctness. We mitigate this limitation by reserving execution-verified correctness checks for benchmark evaluation on CodeForces and LCBv6. Execution-based verification likewise stands in for human evaluation in this domain.

The strongest evidence in this paper is therefore in a code setting with objective execution feedback, and results may not transfer to other code domains such as repository-level workflows. Code editing and agentic software tasks are the planned second instantiation of the regime; no second domain is claimed here. Extending the same engineering discipline to less-verifiable domains (for example, open-ended instruction following) requires stronger verification and accounting standards than we provide here. Our claim is therefore scoped: constrained end-to-end yield engineering can materially improve a brownfield post-training loop in practice. We do not claim universal superiority of this pipeline across models, domains, or mixture designs.

\section*{Ethical Considerations}
\label{sec:ethics}
We follow the ACL Code of Ethics and focus on improving a competitive-programming code model via failure-driven synthetic data generation. We do not use user or personal data.

\noindent\textbf{Data and licensing.} Competitive-programming problems may be subject to platform-specific terms. We treat evaluation artifacts as internal and avoid releasing hidden tests or proprietary logs.

\noindent\textbf{Model misuse.} Improved code-generation capability has dual-use risk. We mitigate by focusing on contest-style tasks and pipeline reliability rather than deployment guidance.

\noindent\textbf{Synthetic data risks.} Synthetic supervision can introduce artifacts. We use judge-based filtering and a deterministic syntax gate to reduce malformed data admission.

\noindent\textbf{Environmental considerations.} Post-training can be compute-intensive; a core objective of our interventions is reducing token and compute cost while maintaining evaluation integrity.

\bibliography{custom}

\begin{thebibliography}{27}
\providecommand{\natexlab}[1]{#1}

\bibitem[{Ahmad et~al.(2025)Ahmad, Narenthiran, Majumdar, Ficek, Jain, Huang,
  Noroozi, and Ginsburg}]{ahmad2025opencodereasoning}
Wasi~Uddin Ahmad, Sean Narenthiran, Somshubra Majumdar, Aleksander Ficek,
  Siddhartha Jain, Jocelyn Huang, Vahid Noroozi, and Boris Ginsburg. 2025.
\newblock \href {https://openreview.net/forum?id=aykM7KUVJZ}
  {{OpenCodeReasoning}: Advancing data distillation for competitive coding}.
\newblock In \emph{Second Conference on Language Modeling}.

\bibitem[{Ben~Allal et~al.(2025)Ben~Allal, Tunstall, Tazi, Bakouch, Beeching,
  Pati{\~n}o, Fourrier, Frere, Lozhkov, Raffel, von Werra, and
  Wolf}]{huggingfacetb2025smolplaybook}
Loubna Ben~Allal, Lewis Tunstall, Nouamane Tazi, Elie Bakouch, Ed~Beeching,
  Carlos~Miguel Pati{\~n}o, Cl{\'e}mentine Fourrier, Thibaud Frere, Anton
  Lozhkov, Colin Raffel, Leandro von Werra, and Thomas Wolf. 2025.
\newblock \href
  {https://huggingface.co/spaces/HuggingFaceTB/smol-training-playbook} {The
  {Smol} training playbook: The secrets to building world-class {LLMs}}.
\newblock Hugging Face.
\newblock Accessed 2026-06-12.

\bibitem[{Brooks(1975)}]{brooks1975mythical}
Frederick~P. Brooks, Jr. 1975.
\newblock \emph{The Mythical Man-Month: Essays on Software Engineering}.
\newblock Addison-Wesley Publishing Company, Reading, Massachusetts.

\bibitem[{Chen et~al.(2021)Chen, Tworek, Jun, Yuan, Pinto, Kaplan, Edwards,
  Burda, Joseph, Brockman, Ray, Puri, Krueger, Petrov, Khlaaf, Sastry, Mishkin,
  Chan, Gray, Ryder, Pavlov, Power, Kaiser, Bavarian, Winter, Tillet,
  Petroski~Such, Cummings, Plappert, Chantzis, Barnes, Herbert-Voss, Guss,
  Nichol, Paino, Tezak, Tang, Babuschkin, Balaji, Jain, Saunders, Hesse, Carr,
  Leike, Achiam, Misra, Morikawa, Radford, Knight, Brundage, Murati, Mayer,
  Welinder, McGrew, Amodei, McCandlish, Sutskever, and
  Zaremba}]{chen2021evaluating}
Mark Chen, Jerry Tworek, Heewoo Jun, Qiming Yuan, Henrique Ponde de~Oliveira
  Pinto, Jared Kaplan, Harri Edwards, Yuri Burda, Nicholas Joseph, Greg
  Brockman, Alex Ray, Raul Puri, Gretchen Krueger, Michael Petrov, Heidy
  Khlaaf, Girish Sastry, Pamela Mishkin, Brooke Chan, Scott Gray, and 39
  others. 2021.
\newblock \href {https://doi.org/10.48550/arXiv.2107.03374} {Evaluating large
  language models trained on code}.
\newblock \emph{arXiv preprint arXiv:2107.03374}.

\bibitem[{Cheng et~al.(2024)Cheng, Lu, Gu, Ke, Liu, Dong, Wang, Tang, and
  Huang}]{cheng2024autodetect}
Jiale Cheng, Yida Lu, Xiaotao Gu, Pei Ke, Xiao Liu, Yuxiao Dong, Hongning Wang,
  Jie Tang, and Minlie Huang. 2024.
\newblock \href {https://doi.org/10.18653/v1/2024.findings-emnlp.397}
  {{AutoDetect}: Towards a unified framework for automated weakness detection
  in large language models}.
\newblock In \emph{Findings of the Association for Computational Linguistics:
  EMNLP 2024}, pages 6786--6803, Miami, Florida, USA. Association for
  Computational Linguistics.

\bibitem[{Fan et~al.(2024)Fan, Pagliardini, and Jaggi}]{fan2024doge}
Simin Fan, Matteo Pagliardini, and Martin Jaggi. 2024.
\newblock \href {https://proceedings.mlr.press/v235/fan24e.html} {{DOGE}:
  Domain reweighting with generalization estimation}.
\newblock In \emph{Proceedings of the 41st International Conference on Machine
  Learning}, volume 235 of \emph{Proceedings of Machine Learning Research},
  pages 12895--12915. PMLR.

\bibitem[{He et~al.(2025)He, Shafique, Kumar, Mackey, and
  Rajani}]{he2025valleycodereasoningscaling}
Muyu He, Muhammad~Ali Shafique, Anand Kumar, Tsach Mackey, and Nazneen Rajani.
  2025.
\newblock \href {https://doi.org/10.48550/arXiv.2510.06101} {The valley of code
  reasoning: Scaling knowledge distillation of large language models}.
\newblock \emph{arXiv preprint arXiv:2510.06101}.

\bibitem[{Jain et~al.(2025)Jain, Han, Gu, Li, Yan, Zhang, Wang, Solar-Lezama,
  Sen, and Stoica}]{jain2024livecodebench}
Naman Jain, King Han, Alex Gu, Wen-Ding Li, Fanjia Yan, Tianjun Zhang, Sida~I.
  Wang, Armando Solar-Lezama, Koushik Sen, and Ion Stoica. 2025.
\newblock \href
  {https://proceedings.iclr.cc/paper_files/paper/2025/hash/94074dd5a072d28ff75a76dabed43767-Abstract-Conference.html}
  {{LiveCodeBench}: Holistic and contamination free evaluation of large
  language models for code}.
\newblock In \emph{International Conference on Learning Representations},
  volume 2025, pages 58791--58831.

\bibitem[{Lehman(1980)}]{lehman1980programs}
Meir~M. Lehman. 1980.
\newblock \href {https://doi.org/10.1109/PROC.1980.11805} {Programs, life
  cycles, and laws of software evolution}.
\newblock \emph{Proceedings of the IEEE}, 68(9):1060--1076.

\bibitem[{Li et~al.(2025{\natexlab{a}})Li, Gao, Wang, Pi, Zhao, Wu, Jiang, Li,
  and Kong}]{li2025reversegen}
Qintong Li, Jiahui Gao, Sheng Wang, Renjie Pi, Xueliang Zhao, Chuan Wu, Xin
  Jiang, Zhenguo Li, and Lingpeng Kong. 2025{\natexlab{a}}.
\newblock \href
  {https://proceedings.iclr.cc/paper_files/paper/2025/hash/1cded4f97cf5f01a284c574110b7e3b9-Abstract-Conference.html}
  {Forewarned is forearmed: Harnessing {LLMs} for data synthesis via
  failure-induced exploration}.
\newblock In \emph{International Conference on Learning Representations},
  volume 2025, pages 10746--10767.

\bibitem[{Li et~al.(2025{\natexlab{b}})Li, Liu, and
  Xing}]{li2025datamixingoptimizationsupervised}
Yuan Li, Zhengzhong Liu, and Eric Xing. 2025{\natexlab{b}}.
\newblock \href {https://proceedings.mlr.press/v267/li25bh.html} {Data mixing
  optimization for supervised fine-tuning of large language models}.
\newblock In \emph{Proceedings of the 42nd International Conference on Machine
  Learning}, volume 267 of \emph{Proceedings of Machine Learning Research},
  pages 35419--35437. PMLR.

\bibitem[{Liang et~al.(2025)Liang, Li, Gong, Wang, Zhang, Shen, Wu, and
  Chen}]{liang2025sws}
Xiao Liang, Zhong-Zhi Li, Yeyun Gong, Yang Wang, Hengyuan Zhang, Yelong Shen,
  Ying~Nian Wu, and Weizhu Chen. 2025.
\newblock \href {https://doi.org/10.52202/085713-1901} {{SwS}: Self-aware
  weakness-driven problem synthesis in reinforcement learning for {LLM}
  reasoning}.
\newblock In \emph{Advances in Neural Information Processing Systems},
  volume~38, pages 56801--56839. Curran Associates, Inc.

\bibitem[{Liu et~al.(2025)Liu, Zheng, Muennighoff, Zeng, Dou, Pang, Jiang, and
  Lin}]{liu2025regmix}
Qian Liu, Xiaosen Zheng, Niklas Muennighoff, Guangtao Zeng, Longxu Dou, Tianyu
  Pang, Jing Jiang, and Min Lin. 2025.
\newblock \href
  {https://proceedings.iclr.cc/paper_files/paper/2025/hash/5f67d864aae6115374fed7beddd119e0-Abstract-Conference.html}
  {{RegMix}: Data mixture as regression for language model pre-training}.
\newblock In \emph{International Conference on Learning Representations},
  volume 2025, pages 38305--38339.

\bibitem[{Madaan et~al.(2024)Madaan, Singh, Schaeffer, Poulton, Koyejo,
  Stenetorp, Narang, and Hupkes}]{madaan2024variance}
Lovish Madaan, Aaditya~K. Singh, Rylan Schaeffer, Andrew Poulton, Sanmi Koyejo,
  Pontus Stenetorp, Sharan Narang, and Dieuwke Hupkes. 2024.
\newblock \href {https://doi.org/10.48550/arXiv.2406.10229} {Quantifying
  variance in evaluation benchmarks}.
\newblock \emph{arXiv preprint arXiv:2406.10229}.

\bibitem[{Miller(2024)}]{miller2024errorbars}
Evan Miller. 2024.
\newblock \href {https://doi.org/10.48550/arXiv.2411.00640} {Adding error bars
  to evals: A statistical approach to language model evaluations}.
\newblock \emph{arXiv preprint arXiv:2411.00640}.

\bibitem[{Rajput et~al.(2025)Rajput, Bonkoungou, Song, Kabore, Olatunji, Klein,
  and Bissyande}]{rajput2025dynamicstability}
Prateek Rajput, Abdoul~Aziz Bonkoungou, Yewei Song, Abdoul~Kader Kabore,
  Iyiola~E. Olatunji, Jacques Klein, and Tegewende Bissyande. 2025.
\newblock \href {https://doi.org/10.48550/arXiv.2511.07463} {Dynamic stability
  of {LLM}-generated code}.
\newblock \emph{arXiv preprint arXiv:2511.07463}.

\bibitem[{Sculley et~al.(2015)Sculley, Holt, Golovin, Davydov, Phillips, Ebner,
  Chaudhary, Young, Crespo, and Dennison}]{sculley2015hidden}
D.~Sculley, Gary Holt, Daniel Golovin, Eugene Davydov, Todd Phillips, Dietmar
  Ebner, Vinay Chaudhary, Michael Young, Jean-Fran\c{c}ois Crespo, and Dan
  Dennison. 2015.
\newblock \href
  {https://proceedings.neurips.cc/paper_files/paper/2015/file/86df7dcfd896fcaf2674f757a2463eba-Paper.pdf}
  {Hidden technical debt in machine learning systems}.
\newblock In \emph{Advances in Neural Information Processing Systems},
  volume~28. Curran Associates, Inc.

\bibitem[{Stan et~al.(2026)Stan, Aflalo, Madasu, Lal, and
  Howard}]{stan2026learningreasoningfailuressynthetic}
Gabriela Ben~Melech Stan, Estelle Aflalo, Avinash Madasu, Vasudev Lal, and
  Phillip Howard. 2026.
\newblock \href {https://doi.org/10.1609/aaai.v40i30.39757} {Learning from
  reasoning failures via synthetic data generation}.
\newblock \emph{Proceedings of the AAAI Conference on Artificial Intelligence},
  40(30):25608--25616.

\bibitem[{{The Microsoft AI Team}(2026)}]{microsoftai2026maithinking}
{The Microsoft AI Team}. 2026.
\newblock \href {https://microsoft.ai/pdf/mai-thinking-1.pdf}
  {{MAI-Thinking-1}: Building a hill-climbing machine}.
\newblock Technical report, Microsoft AI.

\bibitem[{{Tree-sitter contributors}(2026)}]{tree_sitter}
{Tree-sitter contributors}. 2026.
\newblock \href {https://tree-sitter.github.io/tree-sitter/} {Tree-sitter}.
\newblock Project website.
\newblock Accessed 2026-06-12.

\bibitem[{Wang et~al.(2024)Wang, Liu, Shi, Li, Chen, Lu, and
  Yang}]{wang2024inscl}
Yifan Wang, Yafei Liu, Chufan Shi, Haoling Li, Chen Chen, Haonan Lu, and Yujiu
  Yang. 2024.
\newblock \href {https://doi.org/10.18653/v1/2024.naacl-long.37} {{InsCL}: A
  data-efficient continual learning paradigm for fine-tuning large language
  models with instructions}.
\newblock In \emph{Proceedings of the 2024 Conference of the North American
  Chapter of the Association for Computational Linguistics: Human Language
  Technologies (Volume 1: Long Papers)}, pages 663--677, Mexico City, Mexico.
  Association for Computational Linguistics.

\bibitem[{Wang et~al.(2023)Wang, Kordi, Mishra, Liu, Smith, Khashabi, and
  Hajishirzi}]{wang2022selfinstruct}
Yizhong Wang, Yeganeh Kordi, Swaroop Mishra, Alisa Liu, Noah~A. Smith, Daniel
  Khashabi, and Hannaneh Hajishirzi. 2023.
\newblock \href {https://doi.org/10.18653/v1/2023.acl-long.754}
  {{Self-Instruct}: Aligning language models with self-generated instructions}.
\newblock In \emph{Proceedings of the 61st Annual Meeting of the Association
  for Computational Linguistics (Volume 1: Long Papers)}, pages 13484--13508,
  Toronto, Canada. Association for Computational Linguistics.

\bibitem[{Wu et~al.(2025)Wu, Xiang, Wang, Suh, and
  Mittal}]{wu2025thinkingintervention}
Tong Wu, Chong Xiang, Jiachen~T. Wang, G.~Edward Suh, and Prateek Mittal. 2025.
\newblock \href {https://doi.org/10.48550/arXiv.2503.24370} {Effectively
  controlling reasoning models through thinking intervention}.
\newblock \emph{arXiv preprint arXiv:2503.24370}.

\bibitem[{Xie et~al.(2023)Xie, Pham, Dong, Du, Liu, Lu, Liang, Le, Ma, and
  Yu}]{xie2023doremi}
Sang~Michael Xie, Hieu Pham, Xuanyi Dong, Nan Du, Hanxiao Liu, Yifeng Lu,
  Percy~S. Liang, Quoc~V. Le, Tengyu Ma, and Adams~Wei Yu. 2023.
\newblock \href {https://doi.org/10.52202/075280-3059} {{DoReMi}: Optimizing
  data mixtures speeds up language model pretraining}.
\newblock In \emph{Advances in Neural Information Processing Systems},
  volume~36, pages 69798--69818. Curran Associates, Inc.

\bibitem[{Xu et~al.(2026)Xu, Fang, Jiang, Zheng, Xiao, Zhou, Zhao, Zheng, Zhu,
  Tang, Zhao, Luo, Bai, Xu, Su, Wang, Zhao, Qu, and Xu}]{xu2026logicsstem}
Mingyu Xu, Cheng Fang, Keyue Jiang, Yuqian Zheng, Yanghua Xiao, Baojian Zhou,
  Qifang Zhao, Suhang Zheng, Xiuwen Zhu, Jiyang Tang, Yongchi Zhao, Yijia Luo,
  Zhiqi Bai, Yuchi Xu, Wenbo Su, Wei Wang, Bing Zhao, Lin Qu, and Xiaoxiao Xu.
  2026.
\newblock \href {https://doi.org/10.48550/arXiv.2601.01562} {{Logics-STEM}:
  Empowering {LLM} reasoning via failure-driven post-training and document
  knowledge enhancement}.
\newblock \emph{arXiv preprint arXiv:2601.01562}.

\bibitem[{Zhao et~al.(2025)Zhao, Wu, Guan, Gong, and Kong}]{zhao2025promptcot2}
Xueliang Zhao, Wei Wu, Jian Guan, Zhuocheng Gong, and Lingpeng Kong. 2025.
\newblock \href {https://doi.org/10.48550/arXiv.2509.19894} {{PromptCoT} 2.0:
  Scaling prompt synthesis for large language model reasoning}.
\newblock \emph{arXiv preprint arXiv:2509.19894}.

\bibitem[{Zheng et~al.(2026)Zheng, Dong, Liu, Oliva, Yong, Lin, Chen, Wang, and
  Hassan}]{zheng2026whenelo}
Shenyu Zheng, Ximing Dong, Xiaoshuang Liu, Gustavo Oliva, Chong~Chun Yong, Dayi
  Lin, Boyuan Chen, Shaowei Wang, and Ahmed~E. Hassan. 2026.
\newblock \href {https://doi.org/10.48550/arXiv.2602.05891} {When {Elo} lies:
  Hidden biases in {Codeforces}-based evaluation of large language models}.
\newblock \emph{arXiv preprint arXiv:2602.05891}.

\end{thebibliography}

\appendix
\appendixsection{Responsible NLP Checklist (summary)}
\label{sec:responsible-nlp}

This section provides a brief, paper-internal summary of items commonly requested by the ACL Rolling Review ethics policy and ACL submission checklists.
Some checklist items are completed in the submission form (e.g., ARR Responsible NLP Checklist); we include this summary to make key information explicit in the PDF.

\paragraph{Data.}
\begin{itemize}
  \item We do not use user data or personal data; the work uses competitive-programming problems and model-generated synthetic problems.
  \item We do not release proprietary evaluation artifacts (e.g., hidden tests or internal logs).
  \item We run deduplication and decontamination checks ($n$-gram overlap and embedding similarity) against evaluation benchmarks prior to fine-tuning (Section~\ref{sec:setup}).
\end{itemize}

\paragraph{Human subjects.}
\begin{itemize}
  \item No human subjects were recruited, surveyed, or compensated for this study.
\end{itemize}

\paragraph{Compute and efficiency.}
\begin{itemize}
  \item We report a 46.5\% reduction in median teacher reasoning-trace length (mean $-28\%$) under constraint injection with rectified testcases (Figure~\ref{fig:reasoning-length}) as evidence of reduced distillation cost.
\end{itemize}

\paragraph{Risks and mitigations.}
\begin{itemize}
  \item We discuss dual-use risks of improved code generation and mitigate by focusing on contest problems rather than exploit development (see Ethical Considerations).
  \item We discuss risks of LLM-judge bias and preference leakage and mitigate by anchoring acceptance in syntax validity (Tree-sitter).
\end{itemize}

\paragraph{Writing Assistance}
We used an AI-based writing assistant for language editing (clarity/grammar). All technical content, claims, and conclusions were written and verified by the authors.

\appendixsection{Reproducibility: Pipelines, Experiments, and Prompts}
\label{sec:appendix-prompts}
This appendix provides an overview of our methodology, summarizes the concrete experiment instantiation used in this paper, and reproduces the core prompts used in the standard and FDS pipelines.
Our goal is to make the implementation choices (models, filters, and acceptance criteria) explicit so results can be interpreted and replicated.

\subsection{Pipeline definitions (operational order)}
While Figure~\ref{fig:revisited-pipeline} illustrates the closed loop, the operational step order used in our runs is:

\paragraph{Standard pipeline (used for Distill-3K).}
(1) collect evaluation logs on the target benchmark; (2) filter-in test-failing problems; (3) extract concepts; (4) generate synthetic problems; (5) deduplicate; (6) exclude low-quality problems; (7) distill solutions with a fixed attempt cap of 4 teacher queries per synthetic problem (Tree-sitter syntax gate; no execution-based filtering).

\paragraph{FDS pipeline (used for FDS-9K and FDS-3K-Cov).}
Steps (1)--(6) are identical; then (7) testcase rectification: regenerate and correct representative testcases; (8) constraint injection: extract constraint reminders (thinking-intervention sequences); (9) distill solutions with the same fixed attempt cap of 4 teacher queries per synthetic problem (Tree-sitter syntax gate).

\subsection{Benchmarks and Evaluation Protocol}
\label{app:benchmarks}

We evaluate on two execution-verified coding benchmarks using fixed decoding settings. For each condition, we run one fixed checkpoint on each benchmark 16 times, obtaining $n{=}16$ stochastic generations per task. Let $c_q$ denote the number of generations for task $q$ that compile and pass all tests. Following the standard estimator of \citet{chen2021evaluating}, we compute
\begin{equation}
\widehat{\mathrm{pass@}k}(q)
=1-\frac{\binom{n-c_q}{k}}{\binom{n}{k}},
\qquad k\in\{1,3\},
\end{equation}
and report the mean of the task-level estimates. Pass@1 reduces to $c_q/n$ and measures single-generation reliability; pass@3 estimates whether at least one of three independently sampled solutions succeeds, without partitioning the 16 observed generations into arbitrary triplets. We compute 95\% confidence intervals by nonparametric bootstrap over benchmark tasks. For comparisons against Base, each bootstrap replicate uses the same resampled task indices for both conditions, yielding a paired distribution of the mean difference; an interval excluding zero is treated as statistically significant. AIME and MATH report avg@3, the mean across three stochastic regression-suite evaluations; each evaluation score is the fraction of tasks answered correctly. Decoding settings are identical across all conditions and runs.
All training data is decontaminated against benchmark problem statements and hidden tests using $n$-gram overlap and embedding-similarity matching; evaluation uses the benchmarks' hidden tests where applicable. Failure mining uses the separate, stricter protocol defined in Section~\ref{sec:pipeline}.

\paragraph{CodeForces.}
The CodeForces benchmark consists of 65 competitive-programming tasks \citep{zheng2026whenelo}, of which 43 are test-failing and 22 test-passing in the base evaluation logs (``test-passing'' means solved at least once across the $R \times N = 9$ mining samples; the 43 test-failing tasks form the mining pool for $\mathcal{F}$). We evaluate using an offline compile-and-run harness aligned with competitive-programming norms: a generation is correct only if the program compiles and passes all hidden tests. The 16 benchmark evaluations yield $n{=}16$ generations per task; we estimate pass@1 and pass@3 using the estimator above, then average over the 65 tasks.

\paragraph{LiveCodeBench v6 (LCBv6).}
LCBv6 contains 175 tasks and emphasizes longer-context problem specifications \citep{jain2024livecodebench}. The 16 benchmark evaluations yield $n{=}16$ generations per task, from which we report pass@1 and pass@3 using the same execution-based estimator. LCBv6 is never used for failure mining and serves as the held-out transfer benchmark.

\paragraph{Conditions.}
Table~\ref{tab:benchmark_results} reports Base plus additive and mixture-replacement variants of two 3,412-example evaluation patches (Section~\ref{sec:setup}):
\begin{itemize}
  \item \textbf{Base:} the checkpoint after continued training on the 269,198-example rehearsal mixture alone (no new code data).
  \item \textbf{Distill-3K:} standard pipeline without yield engineering, producing 3,412 Tree-sitter-valid synthetic solutions.
  \item \textbf{FDS-3K-Cov:} a 3,412-example coverage-first subset produced with yield engineering (testcase rectification + constraint injection).
\end{itemize}
The evaluations of the final shipped is omitted from the controlled table so the FDS and standard-distillation patches are matched by added example count.

\subsection{Key derived artifacts and counts}
Figure~\ref{fig:revisited-pipeline} annotates the end-to-end sizes for one representative run.
In particular, adding testcase rectification + constraint injection increases successful distillations from 3{,}412 to 9{,}697 solutions (0.84 $\rightarrow$ 2.39 solutions/problem on average), while simultaneously reducing median reasoning length (Figure~\ref{fig:reasoning-length}).


\subsection{Additional plots (space-saving)}
This appendix includes only figures that directly support the paper's core claims (efficiency and benchmark outcomes) and that aid interpretation of the aggregate metrics reported in the main text.
Throughout the operational plots, \textbf{Base} refers to the rehearsal-only \textbf{TuringThinker-7B} condition, and \textbf{Distill-3K} and \textbf{FDS-9K} refer to the corresponding fine-tuned models. These plots retain the full-patch production analysis; Table~\ref{tab:benchmark_results} uses FDS-3K-Cov for the example-count-matched evaluation.

\begin{figure}[h]
  \centering
  \includegraphics[scale=0.32]{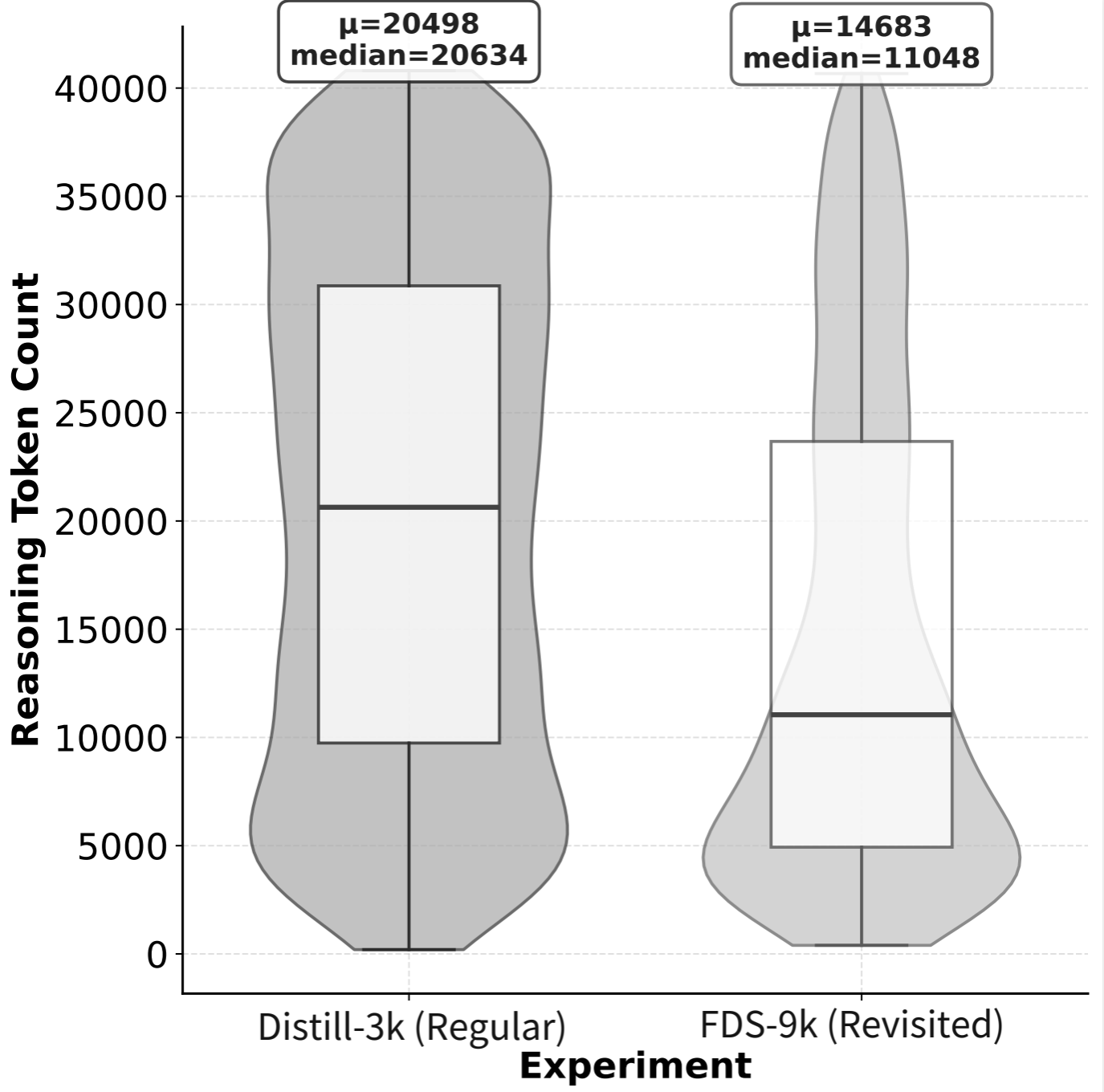}
  \caption{Teacher reasoning length distribution. The FDS interventions shift traces shorter: median trace length falls 46.5\% (mean 28\%).}
  \label{fig:reasoning-length}
\end{figure}

\paragraph{Reasoning-length distribution (cost/latency evidence).}
The main text reports a 46.5\% (median) reduction in teacher reasoning length under constraint injection with rectified testcases. Figure~\ref{fig:reasoning-length} provides a compact visualization of the full distribution shift (median, interquartile range, and outliers), serving as supporting evidence for the cost claim.

\begin{figure}[h]
  \centering
  \includegraphics[scale=0.35]{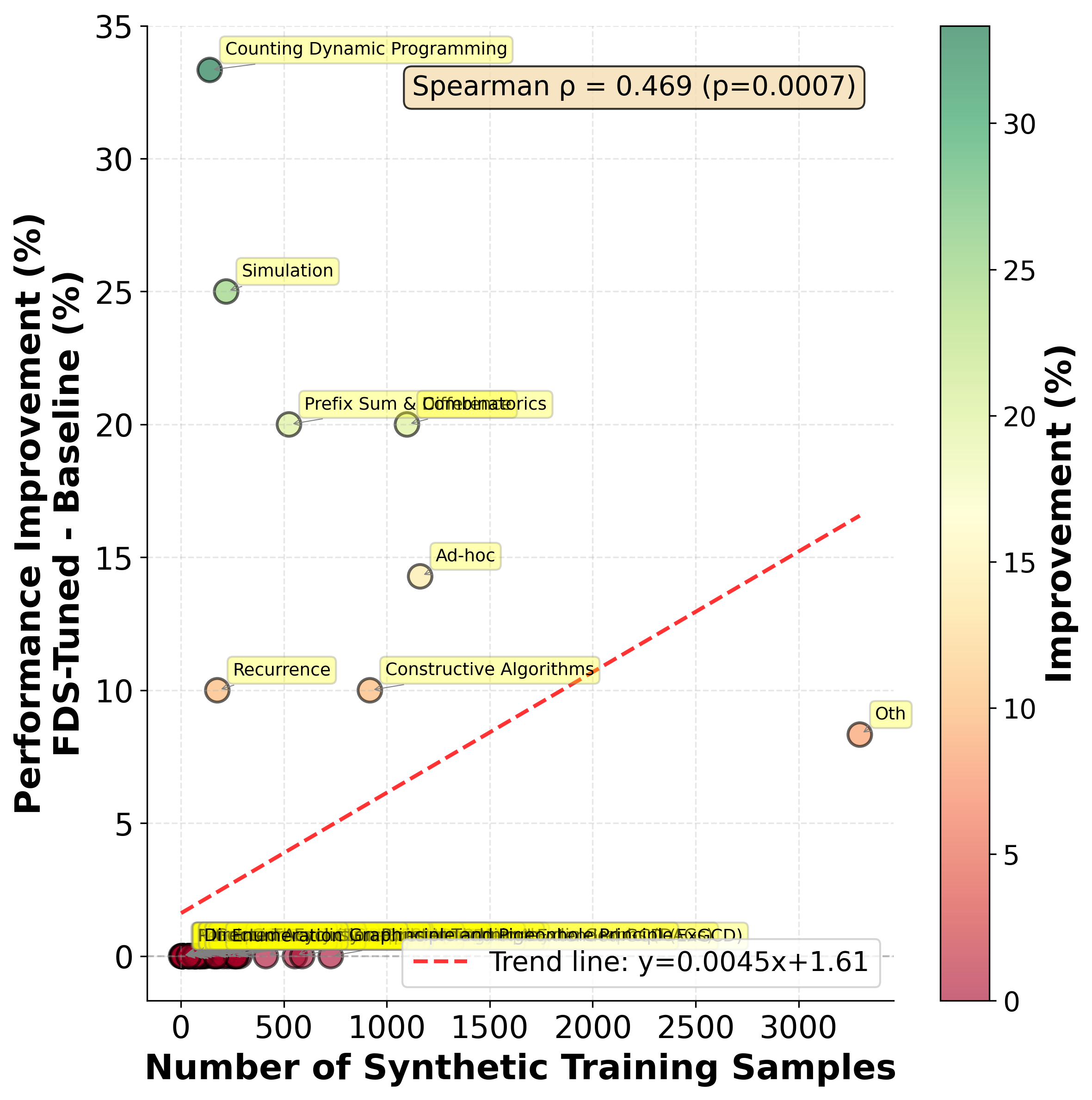}
  \caption{Targeted synthetic volume (per concept) versus downstream improvement. The positive trend (Spearman $\rho = 0.469$) indicates gains concentrate in areas with higher targeted supervision.}
  \label{fig:concept-volume}
\end{figure}

\paragraph{Targeted data volume vs.\ improvement (diagnostic, not the main claim).}
Figure~\ref{fig:concept-volume} visualizes the relationship between \emph{targeted} synthetic data volume (per concept) and downstream improvement on CodeForces. We treat this as a diagnostic for iteration planning (e.g., whether generating more targeted data tends to translate into more robust gains), rather than as primary evidence of effectiveness. Across extracted concepts, downstream CodeForces improvement correlates positively with targeted synthetic volume (Spearman $\rho = 0.469$, $p = 0.0007$, $n = 49$ concepts; Figure~\ref{fig:concept-volume}).
The headline benchmark results are reported in Table~\ref{tab:benchmark_results}.

\makeatletter
\setlength{\@dblfptop}{0pt}
\setlength{\@dblfpsep}{12pt}
\setlength{\@dblfpbot}{0pt plus 1fil}
\makeatother

\begin{figure*}[t!]
\subsection{Prompts (verbatim)}
\centering
\begin{minipage}{0.98\textwidth}
\begin{promptbox}{Concept Extraction Prompt (DeepSeek-V3-1-Terminus)}
\begin{lstlisting}
As an expert in educational assessment, analyze this problem:
Break down and identify {num_concepts} foundational concepts being tested. List these knowledge points that:
- Are core curriculum concepts typically taught in standard courses
- Are precise and measurable (not vague like 'understanding math')
- Are essential building blocks needed to solve this problem
- Represent fundamental principles rather than problem-specific techniques

Think through your analysis step by step, then format your response as a Python code snippet containing a list of {num_concepts} strings, where each string clearly describes one fundamental knowledge point.

# Problem Specification:
{problem}
\end{lstlisting}
\end{promptbox}
\end{minipage}
\caption{Verbatim concept-extraction prompt used with DeepSeek-V3-1-Terminus.}
\label{app:concept-extraction-prompt}
\end{figure*}

\begin{figure*}[t!]
\centering
\begin{minipage}{0.98\textwidth}
\begin{promptbox}{Testcase Regeneration Prompt (DeepSeek-V3.2-Speciale)}
\begin{lstlisting}
As an expert in software testing and quality assurance, generate ONE carefully designed test case for the following programming problem:

=== PROBLEM DESCRIPTION ===
{prompt_text}

=== EVALUATION CRITERIA ===
Your single test case should be selected based on one of these testing priorities:
1. EDGE CASE - A boundary condition that tests limits of the solution
2. COMPLEX SCENARIO - A challenging but valid input that tests multiple aspects
3. ERROR-PRONE CONDITION - An input likely to reveal common implementation mistakes
4. CRITICAL PATH - A test of the most important core functionality

=== REQUIRED OUTPUT FORMAT ===
Provide your response as a JSON object enclosed within code fence markers:

```json
{"input": "<entire input as a single string>", "output": "<expected output as a single string>"}
```

=== FORMAT REQUIREMENTS ===
- The JSON must be enclosed with ```json at the beginning and ``` at the end
- The JSON object must contain ONLY two fields: "input" and "output"
- The input and output must be represented as STRING VALUES, not structured objects
- For multi-line inputs, include newline characters (\n) in the string
- Your response should consist ONLY of the code-fenced JSON object
- Do not include any explanatory text before or after the JSON
- Input/output must exactly match how data would be read/written in a standard input/output stream
\end{lstlisting}
\end{promptbox}
\end{minipage}
\caption{Verbatim testcase-regeneration prompt used with DeepSeek-V3.2-Speciale.}
\end{figure*}

\begin{figure*}[t!]
\centering
\begin{minipage}{0.98\textwidth}
\begin{promptbox}{Constraint Reminder Extraction Prompt (DeepSeek-V3.2-Speciale)}
\begin{lstlisting}
You are a helpful assistant. Your task is to read the user's prompt describing a competitive programming task and extract all hard constraints from it.

# Important Rules
- Do not attempt to solve the competitive programming task itself.
- Do not output anything other than the constraint reminders.

# Output Format
List each constraint as a first-person reminder starting with "I should":
- Use bullet points for multiple constraints
- Be specific
- Use * for bullets (not - or numbers)
\end{lstlisting}
\end{promptbox}
\end{minipage}
\caption{Verbatim constraint-reminder extraction prompt used with DeepSeek-V3.2-Speciale.}
\end{figure*}

\begin{figure*}[t!]
\centering
\begin{minipage}{0.98\textwidth}
\begin{promptbox}{PromptCoT 2.0 Synthetic Problem Generation Prompt}
\begin{lstlisting}
prompt = f"""Given the foundational programming concepts and specified difficulty level, identify connections among these concepts and develop an olympiad-level coding problem that integrates them with appropriate complexity.

Foundational Programming Concepts:
{concept_text}

Difficulty Level: {level}"""
\end{lstlisting}
\end{promptbox}
\end{minipage}
\caption{Verbatim PromptCoT 2.0 synthetic-problem generation prompt.}
\label{app:promptcot-synth}
\end{figure*}

\begin{figure*}[t!]
\centering
\begin{minipage}{0.98\textwidth}
\begin{promptbox}{Judge Prompt (GLM-4.6; verdict = perfect/acceptable/bad)}
\begin{lstlisting}
As a critical expert in educational problem design, evaluate the following problem components:

=== GIVEN MATERIALS ===
1. Problem & Design Rationale: {rationale_and_problem}
   (The rationale describes the author's thinking process and justification in designing this problem)
2. Foundational Concepts: {concept_text}
3. Target Difficulty Level: {level}

=== EVALUATION CRITERIA ===
Rate each criterion as: [Perfect | Acceptable | Bad]
1. FORMAT
- Verify correct implementation of markup tags:
  <!-- BEGIN RATIONALE --> [design thinking process] <!-- END RATIONALE -->
  <!-- BEGIN PROBLEM --> [problem] <!-- END PROBLEM -->
2. FACTUAL ACCURACY
- Check for any incorrect or misleading information in both problem and rationale
- Verify mathematical, scientific, or logical consistency
3. DIFFICULTY ALIGNMENT
- Assess if problem complexity matches the specified difficulty level
- Evaluate if cognitive demands align with target level
4. CONCEPT COVERAGE
- Evaluate how well the problem incorporates the given foundational concepts
- Check for missing concept applications
5. SOLVABILITY
- Verify if the problem has at least one valid solution
- Check if all necessary information for solving is provided

=== RESPONSE FORMAT ===
For each criterion, provide:
1. Rating: [Perfect | Acceptable | Bad]
2. Justification: Clear explanation for the rating

=== FINAL VERDICT ===
After providing all criterion evaluations, conclude your response with:
'Final Judgement: [verdict]'
where verdict must be one of:
- 'perfect' (if both FACTUAL ACCURACY and SOLVABILITY are Perfect, at least two other criteria are Perfect, and no Bad ratings)
- 'acceptable' (if no Bad ratings and doesn't qualify for perfect)
- 'bad' (if ANY Bad ratings)

Note: The 'Final Judgement: [verdict]' line must be the final line of your response.
\end{lstlisting}
\end{promptbox}
\end{minipage}
\caption{Verbatim problem-quality judge prompt used with GLM-4.6.}
\end{figure*}

\end{document}